\documentstyle[preprint,aps,epsfig]{revtex}
\begin{document}
\draft
\title{Langevin dynamics of the glass forming polymer melt: Fluctuations
  around the random phase approximation}
\author{V.G. Rostiashvili$^{(a,b)}$, M. Rehkopf$^{(a)}$ and
        T.A. Vilgis$^{(a)}$}
\address{$^{(a)}$Max-Planck-Institut f\"ur Polymerforschung, Postfach 3148,
D-55021
  Mainz, Germany}
\address{$^{(b)}$ Institute of Chemical Physics, Russian Academy of Science,
142432, Chernogolovka, Moscow region, Russia}
\date{\today}
\maketitle
\begin{abstract}
In this paper the Martin-Siggia-Rose (MSR) functional integral representation
is used for the study of the Langevin dynamics of a polymer melt in terms of
collective variables: mass density and response field density. The resulting
generating functional (GF) takes into account fluctuations around the random
phase approximation (RPA) up to an arbitrary order. The set of equations for
the correlation and response functions is derived. It is generally
shown that for 
cases whenever the fluctuation-dissipation theorem (FDT) holds we arrive at 
equations similar to those derived by Mori-Zwanzig. The case when FDT in the 
glassy phase 
is violated is also qualitatively considered and it is  shown that this
results in a smearing out of the ideal glass transition. The memory
kernel is specified for the ideal glass transition as a sum of all water-melon
diagrams. For the Gaussian chain model the explicit expression for the memory
kernel was obtained and discussed in a qualitative link to the mode-coupling
equation. 
\end{abstract}



                         


\section{Introduction}
Over the last decade the glass transition theory as well as corresponding
experiments were strongly influenced by the mode coupling approach (MCA)
\cite{1,2}. This approach based on the 
Mori-Zwanzig projection formalism \cite{Forster}
specified for the two slow variables: mass density and longitudinal momentum
density. The subsequent projection of the random forces, which are involved in
the memory kernel, onto products of the two densities and the factorization of
the resulting 4-point correlators yields the closed non-linear equation for the
density time correlation function $\phi({\bf k},t)$. The bifurcation analysis
of this equation \cite{1} shows that at some critical values of the coupling
constants and
control parameters (like temperature $T$, density $n$, etc.) the non-vanishing
long time limit $\phi({\bf k},t\rightarrow\infty)=f({\bf k})$ arises. This
indicates the occurance of a non ergodic (glass) state.

A complementary approach based on the non-linear fluctuating hydrodynamic
(NFH) was developed in \cite{3,4,5,6}. The authors start here from a set of
stochastic equations for the mass density and the momentum density. Then by
using the renormalized perturbation theory and one-loop approximation we
derive basically the same equations as in the MCA.

As distinct from the perturbative treatment the numerical solution of these
equations is (with an accuracy of numerical errors) exact. Therefore, a
comparison of the numerical solution with the perturbative results provides an
estimation of the validity of the approximations made in the analytic studies.
The results of these calculations \cite{7,8} show that in spite of the fact
that the time-dependent density correlation functions are slightly stretched the very
important features, such as two-step relaxation regime, can not be
obtained. This regime was reproduced in the numerical study of the Langevin
equations of the system with a free-energy functional of the
Ramakrishnan-Yussouf (RY) form \cite{9}. The RY free energy functional
provides a large number of glassy local minima, but from the results of
ref.\cite{9} it is still not clear whether the observed two-step relaxation
regime arises from non-linearities of density fluctuations in the liquid or
from transitions between  different glassy minima. Moreover, it is not
evident how reliable the results of the NFH are near the main
peak of the static structure factor $S(k)$, i.e. $k\approx k_{0}$. 
In this area the density
fluctuations are rather strong and the correlator of the noise fields in
ref.\cite{8,9} is taken artificially suppressed (the dimensionless parameter
$\lambda=10^{-4}$) otherwise the density could be even negative.

In this situation it is very instructive to start from the opposite, in
some sense, to the NFH limit. In contrast to the NFH the random phase
approximation (RPA) describes the modification of the behavior of
free-particles by effective interactions. The dynamical 
version of the RPA was used mainly for
the description of the dynamic spectrum of simple liquids \cite{10} as well as
for polymer melts and solutions \cite{11^{'},11,12,13}. As have been shown in
ref.\cite{10} the basic defect of the RPA in the context of the glass
transition problem is the absence of a central peak
which is determined by the corresponding slow dynamic \cite{1,2}. 
The reason of this is evident: the RPA completely
ignores intermolecular collisions which only bring the system into a state
of local thermodynamic equilibrium and eventually assure the hydrodynamic
regime. 

In the RPA the particles are free or only weakly interacting. 
Since these effects invariably
dominate at sufficiently short wavelengths, such representation might be
suitable for description at large wavelengths, $k\approx k_{0}$. To assure the
feedback mechanism however, which is responsible for the glass
transition \cite{1,2} or the microphase separation in block copolymers
\cite{14,15,16,17}, approaches beyond RPA are necessary.

For the static case it is such extension which was carried out in
ref.\cite{16,17}. In ref.\cite{RRV} , by using a nonperturbative
Hartree approximation, we have been able to derive a 
generalized Rouse equation for a
tagged chain in a melt. The freezing process
of the  Rouse modes of the test chain was
sequentially considered. In the present paper we emphasizes 
on a systematic way of
taking into account density fluctuations in a 
homopolymer melt with respect to the
dynamic RPA. In doing so the glass transition dynamics will be of our prime
interest.This appears as a fundamental problem in polymer physics. The RPA is
well known in describing several collective phenomena in interacting polymer
systems quite well. The phase behavior (statics and dynamics) of polymer
mixtures, block copolymer melts can be understood very well. The theoretical
description of freezing
processes, however, are certainly beyond the random phase
approximations. Moreover, such
freezing processes cannot be of perturbative nature. Interactions become strong
and dominant on short length scales.
Thus we must use methods that go systematically beyond the classical RPA in
polymer physics. In doing so, the glass transition dynamics in the polymer
melts will be of our prime interest. We must mention one important point
here. So far we restrict ourselves to the cases of low molecular weight
melts. This is to avoid additional complications with reptation dynamics for
melts consisting of chains with a large degree of polymerization. Our point in
this paper is thus to develop a method which allows
to study the freezing of an ensemble of ``Rouse chains''
with a degree of polymerization below the critical molecular weight $N_{c}$.

The paper is organized as follows. In section II the Langevin dynamics of a
homopolymer melt is treated by using the Martin-Siggia-Rose (MSR) generating
functional (GF) method \cite{18,19}. The effective action is represented in
terms of the two collective variables, the mass density and the response field
density. In section III the equations of motion for the 
time-dependent correlation function
and the response function are derived. It is shown that in the regime when
the fluctuation dissipation theorem (FDT) is valid these 
two equations are reduced to
one, having the form of a MCA-equation, which yields under certain conditions
an ideal glass transition. The case when the FDT is violated is also
briefly considered. It is shown that by replacing the usual FDT by the
assumption of a Quasi-FDT (QFDT) this leads to a
smearing out of the ideal glass transition. In section IV the memory kernel
for the Gaussian chain model 
is calculated explicitly and a closed equation for the non-ergodicity
parameter is derived. Finally, section V discusses the main results and
perspectives.

\section{The generating functional of a polymer melt in terms of the
  collective variables}
Consider a homopolymer melt of $M$ chains with the $p-$th chain
configuration at a time moment $t$ characterized by the vector function ${\bf
  R}^{(p)}(s,t)$, where $s$ numerates the segments of the chain, $0\leq s\leq
N$. The simultaneous dynamical evolution of ${\bf R}^{(p)}(s,t)$ is described
by the Langevin equations
\begin{equation}
\xi_{0}\frac{\partial}{\partial t}R_{j}^{(p)}(s,t)+\frac{\delta
  H\{R_{j}^{(p)}\}}{\delta R_{j}^{(p)}}=f_{j}^{(p)}(s,t)\label{La1}
\end{equation}
with the Hamiltonian
\begin{eqnarray}
H\{{\bf R}\}&=&\frac{3T}{2l^{2}}\sum_{p=1}^{M}\int_{0}^{N}ds \left[\frac{\partial
    {\bf R}^{(p)}(s,t)}{\partial
    s}\right]^{2}\nonumber\\
&+&\frac{1}{2}\sum_{p=1}^{M}\sum_{m=1}^{M}\int_{0}^{N}ds\int_{0}^{N}ds'V[{\bf R}^{(p)}(s,t)-{\bf R}^{(m)}(s',t)]\label{La2}
\end{eqnarray} 
and the random force correlator 
\begin{equation}
\left<f_{j}^{(p)}(s,t)f_{i}^{(m)}(s',t')\right>=2T\xi_{0}\delta_{pm}\delta_{ij}\delta(s-s')\delta(t-t')\label{La3}
\end{equation}
where $\xi_{0}$ denotes the bare friction coefficient and from now on the
Boltzmann constant $k_{b}=1$.\\
After using the standard MSR-functional integral representation \cite{18,19}
for the system (\ref{La1}-\ref{La3}), the GF takes the form
\begin{eqnarray}
&Z&\left\{l_{j}^{(p)}(s,t),{\hat
    l}_{j}^{(p)}(s,t)\right\}=\int\prod_{j=1}^{3}\prod_{p=1}^{M}DR_{j}^{(p)}(s,t)D{\hat
  R}_{j}^{(p)}(s,t)\exp{\Bigg \{}A_{0}[R_{j}^{(p)},{\hat
  R}_{j}^{(p)}]\nonumber\\&+&\frac{1}{2}\sum_{p=1}^{M}\sum_{m=1}^{M}\int
dt\int_{0}^{N}ds\:ds'\int\frac{d^{3}k}{(2\pi)^{3}}ik_{j}i{\hat R}_{j}^{(p)}(s,t)V(k)\exp\left\{i{\bf
    k}[{\bf R}^{(p)}(s,t)-{\bf R}^{(m)}(s',t)]\right\}\nonumber\\
&+&\sum_{p=1}^{M}\int
dt\int_{0}^{N}ds\left[R_{j}^{(p)}(s,t)l_{j}^{(p)}(s,t)-i{\hat
    R}_{j}^{(p)}(s,t){\hat l}_{j}^{(p)}(s,t)\right]\Bigg\}\label{GF} 
\end{eqnarray} 
where the MSR-action of the free chain system is given by
\begin{eqnarray}
A_{0}\left\{R_{j}^{(p)},{\hat
  R}_{j}^{(p)}\right\}&=&\sum_{p=1}^{M}\int
dt\int_{0}^{N}ds\Bigg\{T\xi_{0}[i{\hat {\bf R}}^{(p)}(s,t)]^{2}\nonumber\\
&+&i{\hat R}_{j}^{(p)}(s,t)\left[\xi_{0}\frac{\partial}{\partial
    t}R_{j}^{(p)}(s,t)-\varepsilon\frac{\partial^{2}}{\partial
    s^{2}}R_{j}^{(p)}(s,t)\right]\Bigg\}\label{action}
\end{eqnarray}
where $\varepsilon=\frac{3T}{l^{2}}$ is the bare elastic 
modulus of a spring with the
length of a Kuhn segment denoted by $l$, $V(k)$ is the interaction energy of
the chain segments, $l_{j}^{(p)}$ and ${\hat l}_{j}^{(p)}$ are external
fields conjugated to $R_{j}^{(p)}$ and ${\hat R}_{j}^{(p)}$ respectively and
the summation over repeated Cartesian indices is implied. Now
by the same way as in \cite{20} collective variables can be
introduced. As opposed to the statics \cite{14,15,16,17} we need to consider
not only the mass density
\begin{equation}
\rho({\bf r},t)=\sum_{p=1}^{M}\int_{0}^{N}ds\: \delta({\bf r}-{\bf
  R}^{(p)}(s,t))\label{rho}
\end{equation}
but also the longitudinal projection of the response fields
\begin{equation}
\Pi({\bf r},t)=\sum_{p=1}^{M}\int_{0}^{N}ds\: i{\hat R}_{j}^{(p)}(s,t)\nabla_{j}\delta({\bf r}-{\bf
  R}^{(p)}(s,t))\label{Pi}
\end{equation}
where again the summation over repeated Cartesian indices is implied. Then the GF (\ref{GF}) becomes
\begin{eqnarray}
Z\{\cdots\}&=&\int\prod_{p=1}^{M}D{\bf R}^{(p)}(s,t)D{\bf {\hat
  R}}^{(p)}(s,t)D\rho D\Pi\nonumber\\
&\times&\delta\left[\rho({\bf r},t)-\sum_{p=1}^{M}\int ds\:\delta({\bf r}-{\bf
  R}^{(p)}(s,t))\right]\nonumber\\
&\times&\delta\left[\Pi({\bf r},t)-\sum_{p=1}^{M}\int ds\:i{\hat R}_{j}^{(p)}(s,t)\nabla_{j}\delta({\bf r}-{\bf
  R}^{(p)}(s,t))\right]\nonumber\\
&\times&\exp\left\{-\frac{1}{2}\int dt d^{3}r_{1}d^{3}r_{2}\Pi({\bf
    r}_{1},t)\rho({\bf r}_{2},t)V({\bf r}_{1}-{\bf r}_{2})+A_{0}\left\{{\bf
      R}^{(p)},{\hat {\bf R}}^{(p)}\right\}\right\}\label{GF2}
\end{eqnarray}
where the dots imply some source fields which will be specified later.

It is convenient to introduce the 2-dimensional field
\begin{eqnarray}
\rho_{\alpha}(1)={\rho(1) \choose \Pi(1)}\label{2d}
\end{eqnarray}
where $\alpha=0,1$ and $1\equiv ({\bf r}_{1},t_{1})$ is used for 
abbreviation. In terms of the
2-dim. density (\ref{2d}) the GF (\ref{GF2}) takes an especially compact form
\begin{eqnarray}
Z\left\{\psi_{\alpha}\right\}=\int
D\rho_{\alpha}(1)\exp\Bigg\{&-&\frac{1}{4}\rho_{\alpha}({\bar 1})U_{\alpha\beta}({\bar 1}{\bar 2})\rho_{\beta}({\bar 2})+W\{\rho_{\alpha}\}+\rho_{\alpha}({\bar 1})\psi_{\alpha}({\bar 1})\Bigg\}\label{Zkomp}
\end{eqnarray}
with the action of the free system 
\begin{eqnarray}
W\{\rho,\Pi\}&=&\ln\int\prod_{p=1}^{M}D{\bf R}^{(p)}(s,t)D{\hat {\bf
    R}}^{(p)}(s,t)\exp\left\{A_{0}\{{\bf R}^{(p)},{\hat {\bf
      R}}^{(p)}\}\right\}\nonumber\\
&\times&\delta\left[\rho({\bf r},t)-\sum_{p=1}^{M}\int ds\:\delta({\bf r}-{\bf
  R}^{(p)}(s,t))\right]\nonumber\\
&\times&\delta\left[\Pi({\bf r},t)-\sum_{p=1}^{M}\int ds\:i{\hat R}_{j}^{(p)}(s,t)\nabla_{j}\delta({\bf r}-{\bf
  R}^{(p)}(s,t))\right]\label{action2}
\end{eqnarray}
and the 2$\times$2-interaction matrix 
\begin{eqnarray}
U_{\alpha\beta}(1,2)=\left({0\atop V(|{\bf r}_{1}-{\bf r}_{2}|)}{V(|{\bf
    r}_{1}-{\bf r}_{2}|)\atop 0}\right)\label{inter}
\end{eqnarray}
and the $\psi_{\alpha}(1)$ is a source field conjugated to the 2-density
(\ref{2d}). In eq.(\ref{Zkomp}) and below the summation over repeated Greek
indices and integration over variables with bars is implied.\\
The  exact form of the action $W\{\rho_{\alpha}\}$ is not known explicitly,
but can be 
obtained by a functional expansion by assuming that the density fluctuations are not very large and the functional $W\{\rho_{\alpha}\}$ is convex.
\\
The calculation is quite similar to the static case \cite{14,15,16}.
Let us introduce the cumulant GF, the ``free energy'', of the free system
\begin{eqnarray}
F\{\psi_{\alpha}\}&\equiv&\ln Z_{0}\{\psi_{\alpha}\}\nonumber\\
&=&\ln\int D\rho_{\alpha}\:\exp\Big\{W\{\rho_{\alpha}\}+\rho_{\alpha}({\bar
    1})\psi_{\alpha}({\bar 1})\Big\}\label{Fenerg}
\end{eqnarray}
This GF has the expansion
\begin{eqnarray}
F\{\psi_{\alpha}\}&=&F_{\alpha}^{(1)}({\bar 1})\psi_{\alpha}({\bar
  1})+\frac{1}{2!}F_{\alpha\beta}^{(2)}({\bar 1}{\bar 2})\psi_{\alpha}({\bar
  1})\psi_{\beta}({\bar
  2})+\nonumber\\
&+&\frac{1}{3!}F_{\alpha\beta\gamma}^{(3)}({\bar 1}{\bar 2}{\bar 3})\psi_{\alpha}({\bar
  1})\psi_{\beta}({\bar
  2})\psi_{\gamma}({\bar
  3})+\ldots\label{expGf}
\end{eqnarray}
with the {\it free system} cumulant correlators
\begin{eqnarray}
&\:&{F_{\alpha}^{(1)}(1)}=\frac{\delta}{\delta\psi_{\alpha}(1)}F\{\psi_{\alpha}\}\Bigg
|_{\psi=0}\label{Def1}\\
&\:&{F_{\alpha\beta}^{(2)}(1,2)}=\frac{\delta^{2}}{\delta\psi_{\alpha}(1)\delta\psi_{\beta}(2)}F\{\psi_{\alpha}\}\Bigg
|_{\psi=0}\label{Def2}\\
&\:&{F_{\alpha\beta\gamma}^{(3)}(1,2,3)}=\frac{\delta^{3}}{\delta\psi_{\alpha}(1)\delta\psi_{\beta}(2)\delta\psi_{\gamma}(3)}F\{\psi_{\alpha}\}\Bigg
|_{\psi=0}\label{Def3}
\end{eqnarray}
We are searching for the action $W\{\rho_{\alpha}\}$ in the form of an expansion
\begin{eqnarray}
W\{\rho_{\alpha}\}&=&W\left\{\left<\rho_{\alpha}\right>_{0}\right\}+\frac{1}{2!}W_{\alpha\beta}^{(2)}({\bar
  1}{\bar 2})\delta\rho_{\alpha}({\bar 1})\delta\rho_{\beta}({\bar
  2})+\nonumber\\
&+&\frac{1}{3!}W_{\alpha\beta\gamma}^{(3)}({\bar 1}{\bar 2}{\bar 3})\delta\rho_{\alpha}({\bar 1})\delta\rho_{\beta}({\bar
  2})\delta\rho_{\gamma}({\bar
  3})+\ldots\label{exp2}
\end{eqnarray}
where
\begin{eqnarray}
\delta\rho_{\alpha}(1)&=&\rho_{\alpha}(1)-\left<\rho_{\alpha}(1)\right>_{0}\nonumber\\
&=&\rho_{\alpha}(1)-{F_{\alpha}^{(1)}(1)}\label{def2}
\end{eqnarray}
In order to determine the coefficients in the expansion (\ref{exp2}) one
should use the saddle point method when calculating the functional integral
(\ref{Fenerg}). This can be carried out in the same spirit as in refs.~\cite{15,16}. This results in a Legendre transformation
with respect to the extremum field ${\bar \rho}_{\alpha}(1)$:
\begin{eqnarray}
F\{\psi_{\alpha}\}=W\{{\bar \rho}_{\alpha}\}+{\bar \rho}_{\alpha}({\bar
  1})\psi_{\alpha}({\bar 1})\label{Leg1}
\end{eqnarray}
As a result \cite{21} it is found:
\begin{eqnarray}
{\bar \rho}_{\alpha}(1)&=&\frac{\delta F\{\psi_{\alpha}\}}{\delta
  \psi_{\alpha}(1)}\label{Leg2}\\
\psi_{\alpha}(1)&=&-\frac{\delta W\{{\bar \rho}_{\alpha}\}}{\delta {\bar
    \rho}_{\alpha}(1)}\label{Leg3}
\end{eqnarray}
By the use of the expansions (\ref{expGf}) and (\ref{exp2}) in eqs.~(\ref{Leg2},\ref{Leg3}) we
can relate the coefficients $W^{(n)}$ with the free system cumulant
correlators \cite{14,21}
\begin{eqnarray}  
W_{\alpha\beta}^{(2)}(1,2)&=&-\left[\left(F^{(2)}\right)^{-1}\right]_{\alpha\beta}(1,2)\label{Amp1}\\
W_{\alpha\beta\gamma}^{(3)}(1,2,3)&=&F_{\alpha\beta\gamma}^{(3)\:
  amp.}(1,2,3)\label{Amp2}\\
W_{\alpha\beta\gamma\delta}^{(4)}(1,2,3,4)&=&F_{\alpha\beta\gamma\delta}^{(4)
  \:amp.}(1,2,3,4)-\nonumber\\
&-&F_{\alpha\beta{\bar \gamma}}^{(3)
  \:amp.}(1,2,{\bar 3})F_{{\bar \gamma}{\bar {\bar \gamma}}}^{(2)}({\bar 3}{\bar
  {\bar 3}})F_{{\bar {\bar \gamma}}\gamma\delta}^{(3)
  \:amp.}({\bar {\bar 3}},3,4)\label{Amp3}
\end{eqnarray}
where the so called amputated correlators are defined by the relation \cite{21}:
\begin{eqnarray}
F_{\alpha\beta\cdots\delta}^{(n)\:amp.}(1,2,\ldots
,n)=\left(F^{(2)}\right)^{-1}_{\alpha{\bar \alpha}}(1,{\bar
  1})\cdots\left(F^{(2)}\right)^{-1}_{\delta{\bar \delta}}(n,{\bar n})F_{{\bar
    \alpha}\cdots{\bar \delta}}^{(n)\:amp.}({\bar 1}\ldots
,{\bar n})\label{defamp}
\end{eqnarray}
The main result of this section is the GF (\ref{Zkomp}) given now by
\begin{eqnarray}
Z\{\psi_{\alpha}\}=\int
D\rho_{\alpha}\:\exp\Bigg\{&-&\frac{1}{2}\left[U+F^{(2)\:-1}\right]_{\alpha\beta}({\bar
  1},{\bar 2})\delta\rho_{\alpha}({\bar 1})\delta\rho_{\beta}({\bar
  2})\nonumber\\
&+&\frac{1}{3!}W_{\alpha\beta\gamma}^{(3)}({\bar 1}{\bar 2}{\bar 3})\delta\rho_{\alpha}({\bar 1})\delta\rho_{\beta}({\bar
  2})\delta\rho_{\gamma}({\bar
  3})\nonumber\\
&+&\frac{1}{4!}W_{\alpha\beta\gamma\delta}^{(4)}({\bar 1}{\bar 2}{\bar 3}{\bar 4})\delta\rho_{\alpha}({\bar 1})\delta\rho_{\beta}({\bar
  2})\delta\rho_{\gamma}({\bar
  3})\delta\rho_{\delta}({\bar
  4})+\ldots\nonumber\\
&+&\rho_{\alpha}({\bar 1})\psi_{\alpha}({\bar1})\Bigg\}\label{GFN}
\end{eqnarray}
where the coefficients in the effective action are expressed in terms of free
polymer system dynamics and given by eqs.~(\ref{Amp1}-\ref{Amp3},\ref{defamp}). The
relation (\ref{GFN}) is the dynamical generalization of the coarse grained
partition function which was obtained (for a diblock copolymer melt) in the
ref.\cite{15,16}.
\section{The equations of motion for the time correlation and response
  function}
The representation of the GF (\ref{GFN}) is a good starting point for taking into
account fluctuation effects which enable us to go beyond 
the standard RPA methods. Before we proceed in this direction let us
rederive the simple RPA-results for convenience and as a consistency check.
\subsection{RPA-results}
If we restrict ourselves in the expansion (\ref{GFN}) to the quadratic order, then the RPA-correlator and response function $S_{\alpha\beta}(1,2)$ is 
obtained as a $2\times2$-matrix form
\begin{eqnarray}
S_{\alpha\beta}(1,2)=\left[{\hat U}+{\hat
    F}^{(2)\:-1}\right]_{\alpha\beta}^{-1}(1,2)\label{RPA}
\end{eqnarray}
Here ${\hat U}$ is the matrix of interactions and 
given by eq.~(\ref{inter}). The correlation
$2\times2$-matrix of the free system has the form
\begin{eqnarray}
F_{\alpha\beta}^{(2)}(1,2)=\left({F_{00}^{(2)}(1,2)\atop
    F_{10}^{(2)}(1,2)}{F_{01}^{(2)}(1,2)\atop 0}\right)\label{Fdef}
\end{eqnarray}
In eq.~(\ref{Fdef}) off diagonal elements $F_{01}^{(2)}(1,2)$,
$F_{10}^{(2)}(1,2)$ are retarded and advanced susceptibilities respectively;
the diagonal element $F_{00}^{(2)}(1,2)$ is the density correlator. The
relation between them is determined by the FDT in $({\bf k},t)$-representation
\begin{eqnarray}
\beta\frac{\partial}{\partial t}F_{00}^{(2)}({\bf k},t)=F_{01}^{(2)}({\bf
  k},t)-F_{10}^{(2)}({\bf k},t)\label{FDTk}
\end{eqnarray}
or alternatively in $({\bf k},\omega)$-representation
\begin{eqnarray}
\beta i\omega F_{00}^{(2)}({\bf k},\omega)=F_{10}^{(2)}({\bf k},\omega)-F_{01}^{(2)}({\bf
  k},\omega)\label{FDTw}
\end{eqnarray}
where $\beta=1/T$.
The inversion of the $2\times2$-matrix in eq.~(\ref{RPA}) results in the
matrix elements in the $({\bf k},\omega)$-representation
\begin{eqnarray}
S_{00}({\bf k},\omega)&=&\frac{F_{00}^{(2)}({\bf
    k},\omega)}{\left[1+V(k)F_{10}^{(2)}({\bf
      k},\omega)\right]\left[1+V(k)F_{01}^{(2)}({\bf k},\omega)\right]}\label{RPAres1}\\
S_{01}({\bf k},\omega)&=&\frac{F_{01}^{(2)}({\bf
    k},\omega)}{1+V(k)F_{01}^{(2)}({\bf k},\omega)}\label{RPAres2}\\
S_{10}({\bf k},\omega)&=&\frac{F_{10}^{(2)}({\bf
    k},\omega)}{1+V(k)F_{10}^{(2)}({\bf k},\omega)}\label{RPAres3}
\end{eqnarray}
From eqs.~(\ref{RPAres1}-\ref{RPAres3}) we can simply see that the FDT for the
non-interacting  system
(\ref{FDTk}, \ref{FDTw}) assures the validity of FDT in RPA
\begin{eqnarray}
\beta i\omega S_{00}({\bf k},\omega)=S_{10}({\bf k},\omega)-S_{01}({\bf
  k},\omega)\label{SFDTw}
\end{eqnarray}
Eqs.~(\ref{RPAres1}-\ref{RPAres3}) coincide with the classical expressions for RPA-susceptibilities
which have been used in the theory of liquids (see e.g. eq.~(2.7.42) in \cite{10})
and more frequently in the dynamic theory of polymer melts \cite{11,12,13}. 
\subsection{Equations of motion beyond RPA: Role of fluctuations}
In this general case the full renormalized correlator $2\times2-$matrix
$G_{\alpha\beta}(1,2)$ must satisfy the Dyson equation
\begin{eqnarray}
\left[G^{-1}\right]_{\alpha\beta}(1,2)=\left[S^{-1}\right]_{\alpha\beta}(1,2)-\Sigma_{\alpha\beta}(1,2)\label{Dys1}
\end{eqnarray}
where the so called self-energy functional $\Sigma_{\alpha\beta}(1,2)$ contains
only one-line irreducible diagrams (or diagrams which cannot be disconnected
by cutting only one line) \cite{21}. These diagrams are built up from the
vertices given by the effective action (\ref{Zkomp}) and lines with the full
correlator matrix $G_{\alpha\beta}(1,2)$ assigned to them. The diagrams which
are relevant for the glass transition problem will be discussed later. \\
The self-energy matrix has the triangular block structure
\begin{eqnarray}
\Sigma_{\alpha\beta}(1,2)=\left({0 \atop \Sigma_{10}(1,2)}{\Sigma_{01}(1,2)
    \atop \Sigma_{11}(1,2)}\right)\label{selfblock}
\end{eqnarray}
Inversion of the Dyson equation (\ref{Dys1}) yields the result 
\begin{eqnarray}
G_{00}({\bf k},\omega)&=&\frac{S_{00}({\bf k},\omega)S_{01}^{-1}({\bf
    k},\omega)S_{10}^{-1}({\bf k},\omega)+\Sigma_{11}({\bf
    k},\omega)}{\left[S_{01}^{-1}({\bf k},\omega)-\Sigma_{10}({\bf
        k},\omega)\right]\left[S_{10}^{-1}({\bf k},\omega)-\Sigma_{01}({\bf
        k},\omega)\right]}\label{G00}\\
G_{01}({\bf k},\omega)&=&\frac{1}{S_{01}^{-1}({\bf k},\omega)-\Sigma_{10}({\bf
        k},\omega)}\label{G01}\\
G_{10}({\bf k},\omega)&=&\frac{1}{S_{10}^{-1}({\bf k},\omega)-\Sigma_{01}({\bf
        k},\omega)}\label{G10}
\end{eqnarray}
If besides relation (\ref{SFDTw}) FDT is also valid for the full correlator
and response functions
\begin{eqnarray}
\beta i \omega G_{00}({\bf k},\omega)=G_{10}({\bf k},\omega)-G_{01}({\bf
  k},\omega)\label{FDTfull}
\end{eqnarray}
or in time domain
\begin{eqnarray}
\beta\frac{\partial}{\partial t}G_{00}({\bf k},t)=G_{01}({\bf
  k},t)-G_{10}({\bf k},t)\label{FDTfulltime}
\end{eqnarray}
then the exact self-energy obeys
\begin{eqnarray}
\beta i \omega \Sigma_{11}({\bf k},\omega)=\Sigma_{01}({\bf k},\omega)-\Sigma_{10}({\bf
  k},\omega)\label{FDTsigmaw}
\end{eqnarray}
or correspondingly in the time domain
\begin{eqnarray}
\beta\frac{\partial}{\partial t}\Sigma_{11}({\bf k},t)=\Sigma_{10}({\bf k},t)-\Sigma_{01}({\bf
  k},t)\label{FDTsigmat}
\end{eqnarray}
This agrees with the corresponding relations for self-energy given in
\cite{22}. We stress that the FDT for the free system (\ref{FDTw}) as well as in RPA
(\ref{SFDTw}) is always valid. For the full correlator and
response functions of a glass forming system this is not obligatory the case
\cite{23,24,25}.\\
In order to proceed further let us note that for time intervals $t>\tau_{R}$
(where $\tau_{R}$ is the characteristic Rouse time) one can use for the free
system correlator the diffusional approximation \cite{26}
\begin{eqnarray}
F_{00}^{(2)}({\bf k},t)=F_{st}({\bf k})e^{-k^{2}Dt}\label{diffapprox}
\end{eqnarray}
where $D$ is the Rouse diffusion coefficient and $F_{st}({\bf k})$ is the
corresponding static correlator. Then taking into account FDT (\ref{FDTw}) we
have
\begin{eqnarray}
F_{00}^{(2)}({\bf k},\omega)&=&\frac{2k^{2}DF_{st}({\bf
    k})}{\omega^{2}+(k^{2}D)^{2}}\label{F00}\\
F_{01}^{(2)}({\bf k},\omega)&=&-\frac{\beta k^{2}DF_{st}({\bf k})}{-i\omega+k^{2}D}\label{F01}\\
F_{10}^{(2)}({\bf k},\omega)&=&-\frac{\beta k^{2}DF_{st}({\bf
  k})}{i\omega+k^{2}D}\label{F10}
\end{eqnarray}
One should use eqs.~(\ref{F00}-\ref{F10}) in the RPA-result
(\ref{RPAres1}-\ref{RPAres3}), then the expressions for the full correlator and response
functions (\ref{G00}-\ref{G10}), after going back to the time
domain, yields
\begin{eqnarray}
\left[\tau_{c}\frac{\partial}{\partial t}+\chi_{st}^{-1}({\bf
    k})\right]G_{01}({\bf k};t,t')&+&\int_{t'}^{t}\Sigma_{10}({\bf
  k};t,t^{''})G_{01}({\bf k};t^{''},t')dt^{''}=-\delta(t-t')\label{G01t}\\
\left[\tau_{c}\frac{\partial}{\partial t}+\chi_{st}^{-1}({\bf
    k})\right]G_{00}({\bf k};t,t')&+&\int_{-\infty}^{t}\Sigma_{10}({\bf
  k};t,t^{''})G_{00}({\bf k};t^{''},t')dt^{''}\nonumber\\
&+&\int_{-\infty}^{t}\Sigma_{11}({\bf
  k};t,t^{''})G_{10}({\bf
  k};t^{''},t')dt^{''}=-2T\tau_{c}G_{10}(t,t')\label{G00t}
\end{eqnarray}
where the inverse RPA-static susceptibility
\begin{eqnarray}
\chi_{st}^{-1}({\bf k})=\left[\beta F_{st}({\bf
    k})\right]^{-1}-V(k)\label{RPAsus}
\end{eqnarray}
and the bare correlation time
\begin{eqnarray}
\tau_{c}=\frac{1}{\beta k^{2}DF_{st}({\bf k})}\label{cortime}
\end{eqnarray} 
are defined. The initial conditions for the equations (\ref{G01t}) and
(\ref{G00t}) has the form
\begin{eqnarray}
\tau_{c}G_{01}({\bf k};t=t'+0^{+})&=&-1\nonumber\\
G_{01}({\bf k};t=t')&=&0\label{ini1}
\end{eqnarray}
and
\begin{eqnarray}
G_{00}({\bf k};t=t')=G_{st}({\bf k})\label{ini2}
\end{eqnarray}
where $G_{st}({\bf k})$ is the full static
correlator. Eqs.~(\ref{G01t},\ref{G00t}) represent the general result of the
present paper, and below we will discuss the physical aspects of them in more
detail at specific examples.

The resulting equations (\ref{G01t}-\ref{ini2}) are indeed very general. We
made only use of the diffusional approximation for the 
free system correlator (\ref{diffapprox})
and of a causality condition. Qualitatively the same equations of motion
was obtained in the dynamical Hartree approximation for a test chain in a melt
(see eqs.~(29)-(32) in \cite{RRV}) and for a manifold in a random medium (see
eqs.~(B3)-(B5) in \cite{27}). Let us now consider the 
case when besides eqs.~(\ref{G01t}-\ref{ini2}) the FDT
(\ref{FDTfulltime}) for the full correlator and response function holds.


\subsection{Time-homogeneity and FDT for the full matrix
  $G_{\alpha\beta}({\bf k};t)$ hold}
Let us assume in eq.~(\ref{G00t}) $t'=0$ and $t>0$. Then after
differentiation of both sides of eq.~(\ref{G00t}) with respect to the time and
taking into account FDT (\ref{FDTfulltime}) we have
\begin{eqnarray}
\left[\tau_{c}\frac{\partial}{\partial t}+\chi_{st}^{-1}({\bf
    k})\right]G_{01}({\bf k};t)+\int_{0}^{t}\Sigma_{10}({\bf
  k};t-t')G_{01}({\bf k};t')dt'&\:&\nonumber\\
+\int_{-\infty}^{\infty}dt'\left\{\beta\frac{\partial}{\partial
    t}\Sigma_{11}({\bf k};t-t')-\Sigma_{10}({\bf
    k};t-t')\right\}G_{10}({\bf k};t')&=&0\label{red}
\end{eqnarray}
The comparison of the eq.~(\ref{red}) with the eq.~(\ref{G01t}) yields
\begin{eqnarray}
\beta\frac{\partial}{\partial t}\Sigma_{11}({\bf k};t)=\Sigma_{10}({\bf
  k};t)\label{FDTSigma}
\end{eqnarray}
which is again the familiar relation (\ref{FDTsigmat}) for $t>0$. As it should
be the case, one of the eqs.~(\ref{G01t}, \ref{G00t}) is getting redundant
now.

Using the eqs.~(\ref{FDTfulltime}) and (\ref{FDTsigmat}) in eq.~(\ref{G00t}) and
after integration by part we arrive at the result
\begin{eqnarray}
\left[\tau_{c}\frac{\partial}{\partial t}+{\tilde \chi}_{st}^{-1}({\bf
    k})\right]G_{00}({\bf k};t)
+\beta\int_{0}^{t}\Sigma_{11}({\bf k};t-t')\frac{\partial}{\partial
  t'}G_{00}({\bf k};t')dt'=0\label{G00n}
\end{eqnarray}
where
\begin{eqnarray}
{\tilde \chi}_{st}^{-1}({\bf
    k})=\left[\beta F_{st}^{(2)}({\bf k})\right]^{-1}-V({\bf
    k})-\beta\Sigma_{11}({\bf k};t=0)\label{sus}
\end{eqnarray}
The last term in eq.~(\ref{sus}) is the contribution of fluctuations in the
static correlation function. It must be stressed that eq.~(\ref{G00n}) is equivalent to the
Mori-Zwanzig equation, derived by the projection formalism\cite{Forster} and
was specified for the two slow variables, mass density and longitudinal
momentum density, in ref.~\cite{1}.
Performing the Laplace-transformation
\begin{eqnarray}
{\cal L}(\cdots)=\int_{0}^{\infty}dt\ldots\exp(izt)\label{Lap}
\end{eqnarray}
in eq.~(\ref{G00n}) for $\phi({\bf k},z)\equiv G_{00}({\bf k},z)/G_{st}({\bf
    k})$ we get
\begin{eqnarray}
\phi({\bf k},z)=\frac{1}{-iz+{\displaystyle \frac{k^{2}G_{st}^{-1}({\bf
      k})}{\tau_{0}+k^{2}M({\bf k},z)}}}\label{Mori}
\end{eqnarray}
where
\begin{eqnarray}
\tau_{0}&=&\frac{1}{DF_{st}({\bf k})}\\
M({\bf k},z)&=&\beta^{2}\Sigma_{11}({\bf k},z)\label{memoryk}
\end{eqnarray}
and we have used the relation
\begin{eqnarray}
{\tilde \chi}_{st}^{-1}=TG_{st}^{-1}({\bf k})
\end{eqnarray}
The interesting point is that the self-energy matrix element
$\Sigma_{11}$ is proportional to the memory-kernel. Indeed eq.~(\ref{memoryk})
connects the
Mori-Zwanzig technique with the MSR-formalism. The particular form of the
self-energy matrix element $\Sigma_{11}$ for the physical problem defined by
the action given by eq.~(\ref{Zkomp}) will be discussed in the section IV.

It can be expected that a critical temperature $T_{c}$ exists where the
correlator $G_{00}({\bf k},z)$ as well as the memory kernel acquire a pole at
$z=0$. This would show an ergodicity breaking or an ideal glass transition
\cite{1} and will be considered in section IV.
\subsection{The time-homogeneity is valid but FDT is violated}
This case was discussed in the literature \cite{23,27,28}. The more general
case, when the time-homogeneity does not hold any more was also considered
\cite{24,25}.

According the scenario given in \cite{23,27,28} below the temperature $T_{c}$
of the ergodicity breaking the phase space decomposes into regions (ergodic
components). The latter are  separated by high barriers which can not be
crossed at times $t<t^{*}$. The time $t^{*}$ has the physical meaning of a
``trapping'' time. Especially for a system with infinite range 
interactions, we have
$t^{*}\rightarrow \infty$ in the thermodynamic limit. This point had become
most obvious in the case of spin glasses with long range interactions
\cite{23}. In such systems
quenched disorder is present from the beginning and it has been shown
that replica symmetry breaking corresponds to the splitting of the phase
space by infinite high barriers. The present 
dynamical theory on conventional glasses
without quenched disorder suggests a similar
scenario. Disorder develops during cooling and eventually large barriers
develop. Then these barriers cannot be crossed and the phase space is broken
up into accessible parts. Dynamics becomes slow and glassy. Mathematically
this is monitored in the invalidity of the FDT.

To be more precise, we assume that for
our case $t^{*}$ is large but still finite. For the intervals
$t<<t^{*}$ the relaxation occurs within one ergodic component, the
dynamics is ergodic and FDT holds. For $t>>t^{*}$ the system jumps over
the phase space barriers from one component to another and FDT is violated. 
In this case
the total response to an external field consists of two parts. One of them is
determined by the dynamics inside one ergodic component (intracomponent
dynamics) obeying FDT, and the other which appears at $t>t^{*}$ as a result of
crossing the barriers between different components (inter-component dynamics).
 The latter process violates the usual FDT.

With this in mind let us make the following assumption 
\begin{eqnarray}
\beta\left[\frac{\partial}{\partial t}+\gamma
 {\rm sign}(t)\theta(|t|-|t^{*})|\right]G_{00}({\bf k},t)=G_{01}({\bf
  k},t)-G_{10}({\bf k},t)\label{QFDT}  
\end{eqnarray}
which might be the linear version of a more general relation
\begin{eqnarray}
\beta\left[\frac{\partial}{\partial t}+\gamma X\{G_{00}({\bf
    k},t)\}\right]G_{00}({\bf k},t)=G_{01}({\bf k},t)-G_{10}({\bf
  k},t)\label{QFDTg}
\end{eqnarray}
In eqs.~(\ref{QFDT},\ref{QFDTg}) the (phenomenological) parameter $\gamma$ has the dimension of an inverse time
and, as we will see below, has the meaning of a characteristic rate of the
inter-component dynamics ("hopping process" in the nomenclature of
Ref.\cite{2}). $X\{G_{00}\}$ is an arbitrary functional of the correlator $G_{00}({\bf
  k},t)$. In eq.~(\ref{QFDT}) $\theta(\cdots)$ is the $\theta$-function and
${\rm sign}(t)$ keeps the correct transformation under time reversal:
$G_{01}(-t)=G_{10}(t)$. We will call eqs.~(\ref{QFDT},\ref{QFDTg}) after
ref.\cite{27,28} Quasi-FDT (QFDT).

Let us use eq.~(\ref{QFDT}) (at $t>t^{*}>0$) in eq.~(\ref{G00t}) (at $t'=0$
and $t>0$) by acting with $\beta(\partial/\partial t+\gamma)$ on its both
sides. After the same calculations carried out already in section III C we arrive at 
\begin{eqnarray}
\beta\left(\frac{\partial}{\partial t}+\gamma\right)\Sigma_{11}({\bf
  k};t)=\Sigma_{10}({\bf k};t)\label{QFDTs}
\end{eqnarray}
where $t>t^{*}>0$. As before, the eq.~(\ref{QFDTs}) assures that one of the
eqs.~(\ref{G01t},\ref{G00t}) gets redundant, whenever the QFDT (\ref{QFDT}) is
valid.

The substitution of eqs.~(\ref{QFDT}) and (\ref{QFDTs}) into eq.~(\ref{G00t})
yields
\begin{eqnarray}
\left[\tau_{c}\frac{\partial}{\partial t}+{\tilde
    \chi}_{st}^{-1}\right]G_{00}({\bf k};t)&+&\beta\int_{0}^{t}\Sigma_{11}({\bf
  k};t-t')\frac{\partial}{\partial t'}G_{00}({\bf k};t')dt'\nonumber\\
+2\beta\gamma\int_{-\infty}^{-t^{*}}\Sigma_{11}({\bf k};t-t')G_{00}({\bf
  k};t')dt'&+&\beta\gamma\int_{-t^{*}}^{t-t^{*}}\Sigma_{11}({\bf
  k};t-t')G_{00}({\bf k};t')dt'=0\label{QG00}
\end{eqnarray}

Two important limiting cases can be distinguished:
\begin{itemize}
\item If $t\rightarrow\infty$ and $t^*\rightarrow\infty$, but $t/t^*\rightarrow 0$,
then the two last terms in eq.~(\ref{QG00}) can be neglected and we go back to
 eq.~(\ref{G00t}), which contains (under conditions discussed in section IV) an
ideal glass transition. The system could never escape from one ergodic
component (an absolutely confined component in nomenclature of ref.\cite{29}).
\item If $t\rightarrow\infty$ and $t^{*}$ is large but finite, (so that
  $t^{*}/t\rightarrow 0$), then the next to last term in the l.h.s. of
  eq.~(\ref{QG00}) can be neglected. In this case the eq.~(\ref{QG00}) takes
  the form
\begin{eqnarray}
\left[\tau_{c}\frac{\partial}{\partial t}+{\tilde
    \chi}_{st}^{-1}\right]G_{00}({\bf k};t)+\beta\int_{0}^{t}\Sigma_{11}({\bf
  k};t-t')\left(\frac{\partial}{\partial t'}+\gamma\right)G_{00}({\bf
  k};t')dt'\label{hoppt}
\end{eqnarray}
The Laplace transformation of eq.~(\ref{hoppt}) gives the result
\begin{eqnarray}
\phi({\bf k},z)=\frac{1}{-iz+\gamma+{\displaystyle
    \frac{k^{2}{\bar\chi}_{st}^{-1}(k)/T}{\tau^{0}+k^{2}M({\bf
        k},z)}}}\label{hoppz}
\end{eqnarray}
with
\begin{eqnarray}
{\bar\chi}_{st}^{-1}={\tilde \chi}_{st}^{-1}-\gamma\tau_{c}\label{chibar}
\end{eqnarray}
In contrast to the ideal glass transition case at $z\rightarrow 0$ the
kernel $M({\bf k},z)$ is large but finite. Then , according
eq.~(\ref{hoppz}) at $z\rightarrow 0$ the behavior of $\phi({\bf
  k},z)$ is determined mainly by a simple pole at $iz=\gamma$. Instead of
going to a plateau the correlator $\phi({\bf k},t)$ decays with the
characteristic rate $\gamma$. This is a result of the inter-component dynamics.

\end{itemize}
\section{The explicit form for the memory kernel $M({\bf k},z)$. The
  mode-coupling approximation.}
For the ideal glass transition problem \cite{1,2} the correlator $G_{00}({\bf
  k},z)$ as well as the memory kernel $M({\bf k},z)$ acquire a pole at $z=0$
at the critical temperature $T_{c}$. One can easily see that for the effective
action, given by eq.~(\ref{Zkomp}), such contribution come from the sum of all
"water-melon" diagrams which are represented in Fig.1. Each line denotes the
full matrix $G_{\alpha\beta}({\bf k},z)$ and a vertex with $m$ legs denotes the
bare vertex function $W^{(m)}_{\alpha\beta\ldots\gamma}({\bf k}_{1},z_{1};{\bf
  k}_{2},z_{2};\ldots{\bf k}_{m},z_{m})$. In the MCA all vertices
renormalization is neglected \cite{30}.\\
Another type of diagram, the "tadpole" diagrams, which are shown in Fig.2,
appears in the context of a Hartree-approximation \cite{27,28,31}. On the
other hand the contributions of these diagrams remain
finite at $z\rightarrow 0$ and because of this they are not relevant for
the ideal glass transition problem. On the contrary, these diagrams are
essential in the context of e.g. fluctuation effects in the theory of
microphase separation in block copolymers \cite{16,17}. It was shown there,
that fluctuations change the order of the phase transition from two (mean
field) to one (upon renormalization). In this research field the corresponding
procedure has been
terminated as Brazovskii renormalization.

The explicit expression for the arbitrary vertex function
$W^{(n)}_{\alpha\beta\ldots\gamma}(1,2,\ldots n)$ is not known in detail. 
That is why we can restrict
ourselves only to the first diagram given in Fig.1, 
which corresponds to the one-loop
approximation. As a result we have  
\begin{eqnarray}
\Sigma_{11}({\bf
  k};z)=2(\frac{1}{3!})^{2}\int\frac{d^{4}q}{(2\pi)^{4}}W_{1{\bar \alpha}{\bar
    \beta}}^{(3)}(k,k-q,q)G_{{\bar\alpha}{\bar\gamma}}(k,k-q)G_{{\bar\beta}{\bar\delta}}(q)W_{{\bar\gamma}{\bar \delta}1}^{(3)}(q,k-q,k)\label{S11}
\end{eqnarray}
where the short hand notation, $k\equiv({\bf k},z)$ and $q\equiv({\bf q},s)$,
was used and the expression for the vertex function has the specific form
\begin{eqnarray}
W_{\alpha\beta\gamma}^{(3)}(1,2,3)=F_{{\bar\alpha}{\bar\beta}{\bar\gamma}}^{(3)}({\bar
  1},{\bar 2},{\bar 3})F_{{\bar\alpha}\alpha}^{(2)-1}({\bar
  1},1)F_{{\bar\beta}\beta}^{(2)-1}({\bar
  2},2)F_{{\bar\gamma}\gamma}^{(2)-1}({\bar 3},3)\label{W3}
\end{eqnarray}
In general there are nine terms in the sum (\ref{S11}), but only one of them,
which does not include a response function has the important $1/z$-singularity at
$z\rightarrow 0$. This leads to the expression
\begin{eqnarray}
\Sigma_{11}({\bf
  k},z)=2(\frac{1}{3!})^{2}\int\frac{d^{4}q}{(2\pi)^{4}}W_{100}^{(3)}(k,k-q,q)G_{00}(k-q)G_{00}(q)W_{001}^{(3)}(q,k-q,k)\label{S11n}
\end{eqnarray}
where
\begin{eqnarray}
W_{100}^{(3)}(k,k-q,q)=F_{011}^{(3)}(k,k-q,q)(F^{(2)-1})_{01}(k)(F^{(2)-1})_{10}(k-q)(F^{(2)-1})_{10}(q)\label{W100}
\end{eqnarray}
and
\begin{eqnarray}
W_{001}^{(3)}(q,k-q,k)=F_{110}^{(3)}(q,k-q,k)(F^{(2)-1})_{10}(q)(F^{(2)-1})_{10}(k-q)(F^{(2)-1})_{01}(k)\label{W001}
\end{eqnarray}
In eq.~(\ref{S11n}) the integrals over Laplace-frequency $s$ is taken along
the straight line in the complex $s$-plane above all singularities of the
integrand. Since a pole of $G_{00}({\bf q},s)$ at $s$=0 
predetermines the $1/z$-behavior of the whole integral, we are able to consider the
vertex functions $W_{100}^{(3)}$ and $W_{001}^{(3)}$ only in the static limit,
$s\rightarrow 0$ and $z\rightarrow 0$. Below, we give these limits for the $2$-
and $3$-point response functions.\\
The Laplace transformation of the 3-point response function, which appears in
eq.~(\ref{W100}), is given by eq.~(\ref{A4}) in the Appendix. As a result its
static limit reads as
\begin{eqnarray}
\lim_{z_{2},z_{3}\rightarrow
  0}F_{011}^{(3)}({\bf k}_{2,}z_{2};{\bf k}_{3},z_{3})=\beta^{2}F_{st}^{(3)}({\bf k_{2}},{\bf
  k_{3}})\label{Fst1}
\end{eqnarray}
The static limit of the 3-point response function, which appears in
eq.~(\ref{W001}) has the same form
\begin{eqnarray}
\lim_{z_{2},z_{3}\rightarrow
  0}F_{110}^{(3)}(z_{1},z_{2})=\beta^{2}F_{st}^{(3)}({\bf k_{1}},{\bf
  k_{2}})\label{Fst2}
\end{eqnarray}
The static limit for the 2-point response functions is given by the relation
\begin{eqnarray}
\lim_{z\rightarrow 0}F_{01}^{(2)}({\bf k},z)=\lim_{z\rightarrow
  0}F_{10}^{(2)}({\bf k},z)=-\beta F_{st}({\bf k})\label{Fst3}
\end{eqnarray}
Explicitly the expressions for $F_{st}^{(3)}$ and $F_{st}^{(2)}$ are given by
\cite{33}:
\begin{eqnarray}
F_{st}^{(3)}({\bf K}_{1},{\bf
  K}_{2})&=&-\frac{1}{2}N^{2}\rho_{0}\nonumber\\
&\times&\left\{\frac{J_{2}({\bf
      K}_{1}^{2})-J_{2}({\bf K}_{2}^{2})}{{\bf K}_{1}^{2}-{\bf K}_{2}^{2}}+\frac{J_{2}({\bf
      K}_{2}^{2})-J_{2}({\bf K}_{3}^{2})}{{\bf K}_{2}^{2}-{\bf K}_{3}^{2}}+\frac{J_{2}({\bf
      K}_{3}^{2})-J_{2}({\bf K}_{1}^{2})}{{\bf K}_{3}^{2}-{\bf K}_{1}^{2}}\right\}\label{49a}
\end{eqnarray}
with
\begin{eqnarray}
{\bf K}_{1}+{\bf K}_{2}+{\bf K}_{3}=0\label{trans}
\end{eqnarray}
and
\begin{eqnarray}
F_{st}^{(2)}({\bf K})=N\rho_{0}J_{2}({\bf K}^{2})\label{49b}
\end{eqnarray}
where ${\bf K}={\bf k}l\sqrt{N/6}$ and $\rho_{0}=cN$ is the average segments
concentration. The function
\begin{eqnarray}
J_{2}(x)=2\frac{e^{-x}-1+x}{x^{2}}\label{49c}
\end{eqnarray}
is known as the De-bye function in polymer physics and is usually approximated
well by the more simple Pad\'e  expression \cite{26,33}
\begin{eqnarray}
J_{2}(x)\approx\frac{1}{1+\frac{1}{2}x}\label{49d}
\end{eqnarray}
By making use of eqs.~(\ref{49a}-\ref{49d}) the expression for the vertices
(\ref{W100},\ref{W001}) in the static limit becomes
\begin{eqnarray}
W_{st}^{(3)}({\bf K}_{1},{\bf K}_{2},{\bf
  K}_{3})=-\frac{1}{4\beta\rho_{0}^{2}N}\left[3+\frac{1}{2}({\bf
    K}_{1}^{2}+{\bf K}_{2}^{2}+{\bf K}_{3}^{2})\right]\label{50}
\end{eqnarray}
where ${\bf K}_{i}^{2}={\bf k}_{i}^{2}l^{2}N/6$. Then for the memory kernel
(\ref{memoryk}) we derive
\begin{eqnarray}
M({\bf k},z)&=&2\left(\frac{1}{4!}\right)^{2}\frac{1}{\rho_{0}^{4}N^{2}}
\int\frac{d^{3}qds}{(2\pi)^{4}}\left[3+\frac{Nl^{2}}{6}(k^{2}+q^{2}+{\bf
    k}{\bf q})\right]^{2}G_{st}(-{\bf k}-{\bf q})G_{st}({\bf q})\nonumber\\
&\times&\phi(-{\bf
  k}-{\bf q},-z-s)\phi({\bf q},s)\label{51}
\end{eqnarray}
The integral over $q$ is mainly determined by the strong peak of $G_{st}({\bf
  q})$ at $q=q_{0}=1/\sigma$, where $\sigma$ is the bead diameter in the
spring-bead model for the chains in the melt. In that case the second term in
the brackets dominates and $N$-dependence for long chains is cancelled, as it should be.\\
In the glass state the correlator $\phi({\bf k},z)$ has the form
\begin{eqnarray}
\lim_{z\rightarrow 0}\phi({\bf k},z)=\frac{f({\bf k})}{-iz}\label{52}
\end{eqnarray}
where $f({\bf k})$ is the non-ergodicity parameter. From eqs.~{\ref{Mori}} and
(\ref{51}) one can easily see that the function $f({\bf k})$ satisfies the
equation
\begin{eqnarray}
\frac{f({\bf k})}{1-f({\bf
    k})}&=&2\left(\frac{1}{4!6}\right)^{2}\left(\frac{l}{\rho_{0}}\right)^{4}G_{st}({\bf k})\nonumber\\
&\:&\int\frac{d^{3}qds}{(2\pi)^{4}}[k^{2}+q^{2}+{\bf
    k}{\bf q}]^{2}G_{st}(-{\bf k}-{\bf q})G_{st}({\bf q})f(-{\bf k}-{\bf
    q})f({\bf q})\label{53}
\end{eqnarray}
which qualitatively corresponds to the result of the conventional MCT (see
e.g. eq.~(3.37) in ref.\cite{1}). The factor $l^{4}$ in front of the
mode-coupling integral (\ref{53}) indicates that the relevant length scale
for the glass area is indeed the Kuhn segment length. 
This confirms statements, that in polymer melts, where a large variety and
wide range of
internal degrees of freedom dominate the physical behavior the glass transition
is indeed ruled on local scales, i.e., the range of the nearest neighbor
interactions. 
Moreover, the present
model suggests, that the glass area is larger for stiffer
chains, since $l$ becomes larger.
This is qualitatively correct but needs a more
detailed investigation.

\section{Summary and discussion}
In this paper we have shown that the MSR-functional integral representation is
very convenient for the treatment 
of the Langevin dynamics 
of the polymer melt
in terms of the 
collective variables: the mass density and the response field density. The
expansion of the free energy and the Legendre transformation technique, which
was given in the ref.\cite{14,15,16} for the static case, was extended here
for the dynamics. As a result we have derived the dynamical GF (\ref{Zkomp})
with the action that allow results 
beyond the RPA up to arbitrary order of the density
and/or response fields density.

It was shown that the
GF is a good starting point for the derivation of the general dynamical
equations. This is the set of equations (\ref{G01t},\ref{G00t}) 
for the correlation
and response functions, with the free part determined by RPA. For the
particular case when the time-homogeneity and FDT are satisfied these   
two equations reduce to the one eq.~(\ref{G00n}).

It is obvious that eq.~(\ref{G00n}) is equivalent to the Mori-Zwanzig equation
with the memory kernel given by the matrix element $\Sigma_{11}({\bf
  k},t)$. In the framework of the Mori-Zwanzig formalism and MCA the glass
transition problem was extensively discussed in \cite{1,2}. In our approach
these results are qualitatively restored if the ``water-melon'' diagrams for
$\Sigma_{11}({\bf k},z)$ (see Fig.1) are summarized. The results of the
vertices renormalization could be in principle also investigated \cite{34}.

The situation when FDT is violated is much more obscure in spite of the
extensive discussion \cite{23,24,25,27,28}. It could be  
guessed that the physical reason
for this (at least at a finite $t^{*}$) is an intercomponent dynamics as it
was discussed in section IV D. We 
have made a simple but plausible assumption about the form of
QFDT (\ref{QFDT}). When this assumption was made in the corresponding
model calculation we had shown that a sharp (or ideal) glass transition
is smeared out by this intercomponent dynamics with a characteristic rate
$\gamma$ and the density correlator takes the form (\ref{hoppz}).

For the ideal glass transition case we have calculated the expression for the
memory kernel (\ref{51}) by using the explicit forms for the free system's
static correlators. As a result the equation (\ref{53}) for the non-ergodicity
parameter $f({\bf k})$ is qualitatively the same like in MCA. The physical
picture we had derived here is most interesting. We have indeed shown that the
violation of the fluctuation dissipation theorem yields a similar picture as
in systems where quenched disorder is present from the beginning. For
example in spin glasses it was shown, that the presence of disorder and
frustration yields a glassy phase. Moreover it could be shown that the case of
replica symmetry breaking is responsible for the disordered nature of the
phase space, especially for the large barriers dividing the phase space. Thus
the entire phases space is no longer accessible for the system. This
corresponds e.g.  to glassy dynamics. In structural glasses such as polymer
melts which form  easily glasses, quenched disorder is not present. In the
melt phase equilibrium dynamics determines the structural properties,
even though the dynamics is very slow mainly ruled by the molecular
weight. Here we must start from equations that take into account all
interactions. With the formalism presented in the sections above we succeeded
to derive a similar picture on totally different physical grounds. We had
shown that glassy phases could appear upon violation of the FDT. At a simple
model computation we had presented arguments, that a characteristic time $t^{*}$
can be related to barrier heights that eventually can be crossed. This
corresponds to the breaking of the phase space into different regions that are
separated by barriers, which have their origin in the interactions. In fact,
these barriers correspond to the development of a multi valley structure of the
phase space. We will come back to this point in later publications.

The present general formalism allows important generalizations for interacting
systems. For example
we will treat  homopolymer blends and  copolymer melts which 
tremendously enriches
the picture of the dynamical behavior. The dynamics of copolymers was studied
above a microphase separation temperature $T_{ms}$ in RPA \cite{12,13} and
below $T_{ms}$ by the numerical solution of the Ginzburg-Landau equation
\cite{35}. One can easily derive a copolymer counterpart of the dynamical GF
(\ref{Zkomp}). Then the Hartree approximation, or summation of all tadpole
diagrams shown in Fig.2 (in the same manner as for the static \cite{16,17})
gives a direct way to obtain a closed dynamical equation for the
composition-composition correlation function.
Moreover we have now the possibility to study
the glass transition in blends and
copolymer melts. These problems are of wide experimental interest. Imagine for
example that one component of the blend or one species of the copolymer melt
freezes out during cooling. The striking problem is then to discuss the
interplay between freezing and phase - or microphase separation. The dynamics
of one species
becomes very slow and some 
parts of the systems become eventually immobile at 
certain correlation length, corresponding to the distance of the (ideal)
critical point. These problems are also under current investigation.

\section{Acknowledgements}
The authors thank the Deutsche Forschungsgemeinschaft (DFG), the
Sonderforschungsbereich SFB 262, and the Bundesministerium f\"ur Bldung und
Forschung (BMBF) for finacial support of the work.

\begin{appendix}
\section{The 3-point response function and its Laplace transformation}
The 3-point response function which appears in eq.~(\ref{W100}) is determined
by
\begin{eqnarray}
&\:&F_{011}^{(3)}({\bf k}_{1},t_{1};{\bf k}_{2},t_{2};{\bf
  k}_{3},t_{3})=(ik_{2})_{j}(ik_{3})_{l}\sum_{p_{1},p_{2},p_{3}=1}^{M}\int_{0}^{N}ds_{1}ds_{2}ds_{3}\label{A1}\\
&\times&\left<i{\hat R}_{j}^{(p_{2})}(s_{2},t_{2})i{\hat
    R}_{l}^{(p_{3})}(s_{3},t_{3})\exp\left\{i{\bf k}_{1}{\bf
      R}^{(p_{1})}(s_{1},t_{1})+i{\bf k}_{2}{\bf
      R}^{(p_{2})}(s_{2},t_{2})+i{\bf k}_{3}{\bf R}^{(p_{3})}(s_{3},t_{3})\right\}\right>_{0}\nonumber
\end{eqnarray}
where $<\cdots>_{0}$ stands for the averaging with the action of the free chain
system. By making use of the Nonlinear-FDT (NFDT) rule (see eq.~(2.31) in
ref.\cite{32}) 
\begin{eqnarray}
-i{\hat R}_{j}(s,t)\longrightarrow\beta\frac{\partial}{\partial
  t}R_{j}(s,t)\label{A2}
\end{eqnarray}
we will come to the following relation
\begin{eqnarray}
F_{011}^{(3)}({\bf k}_{1},t_{1};{\bf k}_{2},t_{2};{\bf
  k}_{3},t_{3})=\beta^{2}\frac{\partial^{2}}{\partial t_{2}\partial t_{3}}F_{000}^{(3)}({\bf k}_{1},t_{1};{\bf k}_{2},t_{2};{\bf
  k}_{3},t_{3})\label{45b}
\end{eqnarray}
with the causality condition $t_{1}>\{t_{2},t_{3}\}$. In the same way we get
\begin{eqnarray}
F_{110}^{(3)}({\bf k}_{1},t_{1};{\bf k}_{2},t_{2};{\bf
  k}_{3},t_{3})=\beta^{2}\frac{\partial^{2}}{\partial t_{1}\partial t_{2}}F_{000}^{(3)}({\bf k}_{1},t_{1};{\bf k}_{2},t_{2};{\bf
  k}_{3},t_{3})\label{45a}
\end{eqnarray}
with the causality condition $t_{3}>\{t_{1},t_{2}\}$.\\
Let us define the Laplace-transformation of the 3-point response function
\begin{eqnarray}
F_{011}^{(3)}(z_{1},z_{2},z_{3})&=&\int_{-\infty}^{\infty}dt_{1}\int_{-\infty}^{t_{1}}dt_{2}\int_{-\infty}^{t_{1}}dt_{3}F_{011}^{(3)}(t_{1},t_{2},t_{3})\exp\{iz_{1}t_{1}+iz_{2}t_{2}+iz_{3}t_{3}\}\nonumber\\
&=&2\pi\delta(z_{1}+z_{2}+z_{3})\int_{-\infty}^{0}dt_{21}\int_{-\infty}^{0}dt_{31}F_{011}^{(3)}(t_{21},t_{31})\exp\{iz_{2}t_{21}+iz_{3}t_{31}\}\nonumber\\
&=&2\pi\delta(z_{1}+z_{2}+z_{3})F_{011}^{(3)}(z_{2},z_{3})\label{A3}
\end{eqnarray} 
where we have used the causality condition $t_{1}>\{t_{2},t_{3}\}$ and the time
translational invariance.\\
By making use of the NFDT (\ref{45a})
and after integrations by parts we have
\begin{eqnarray}
F_{011}^{(3)}({\bf k}_{2},z_{2};{\bf
  k}_{3},z_{3})&=&\beta^{2}\Bigg\{F_{st}^{(3)}({\bf k}_{2},{\bf
  k}_{3})-iz_{2}F_{000}^{(3)}({\bf k}_{2},z_{2};{\bf
  k}_{3},t_{31}=0)\nonumber\\
&-&
iz_{3}F_{000}^{(3)}({\bf k}_{2},t_{21}=0;{\bf k}_{3},z_{3})+(iz_{2})(iz_{3})F_{000}^{(3)}({\bf k}_{2},z_{2};{\bf
  k}_{3},z_{3})\Bigg\}\label{A4}
\end{eqnarray}
where
\begin{eqnarray}
F_{000}^{(3)}({\bf k}_{2},z_{2};{\bf
  k}_{3},z_{3})&=&\int_{-\infty}^{0}dt_{31}\int_{-\infty}^{0}dt_{21}F_{000}^{(3)}({\bf k}_{2},t_{21};{\bf k}_{3},t_{31})\exp\{iz_{2}t_{21}+iz_{3}t_{31}\}
\end{eqnarray}
and
\begin{eqnarray}
F_{000}^{(3)}({\bf k}_{2},z_{2};{\bf
  k}_{3},t_{31}=0)&=&\int_{-\infty}^{0}dt_{21}F_{000}^{(3)}({\bf
  k}_{2},t_{21};{\bf k}_{3},t_{31}=0)\exp\{iz_{2}t_{21}\}
\end{eqnarray}
and $F_{st}^{(3)}({\bf k}_{2},{\bf k}_{3})$ is the static 3-point density
correlator. From eq.~(\ref{A4}) we immediately obtain the static limit
(\ref{Fst1}).
\end{appendix}

\newpage
\begin{figure}
\epsfig{file=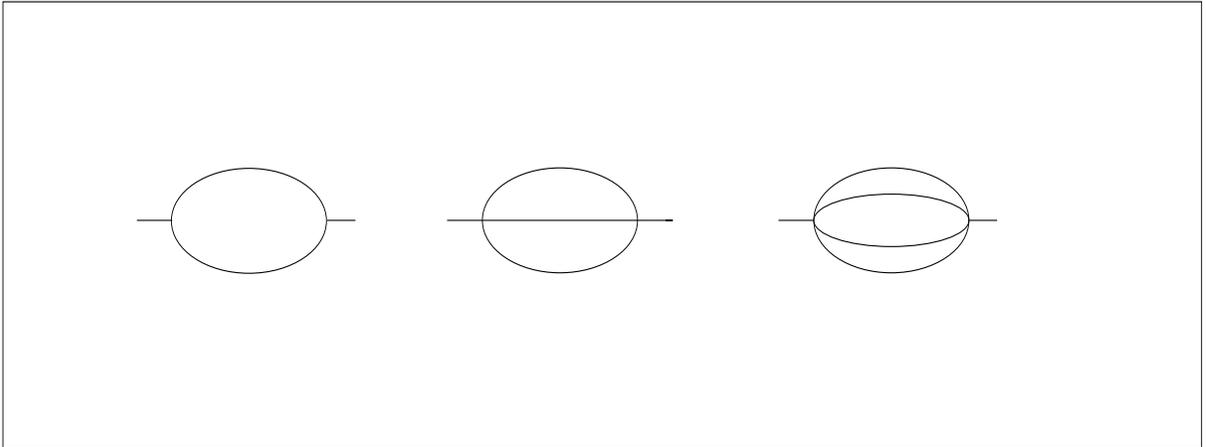, width=16cm, height=6cm}
\caption{Diagrammatic representation of the self-energy $\Sigma_{11}({\bf
    k},z)$ in MCA, which has a simple pole at $z=0$, i.e. is relevant for the
ideal glass transition. The vertices are bare: $W^{(3)},W^{(4)},\ldots$,
i.e. this approximation neglects all vertex renormalization.}
\end{figure}
\newpage
\begin{figure}
\epsfig{file=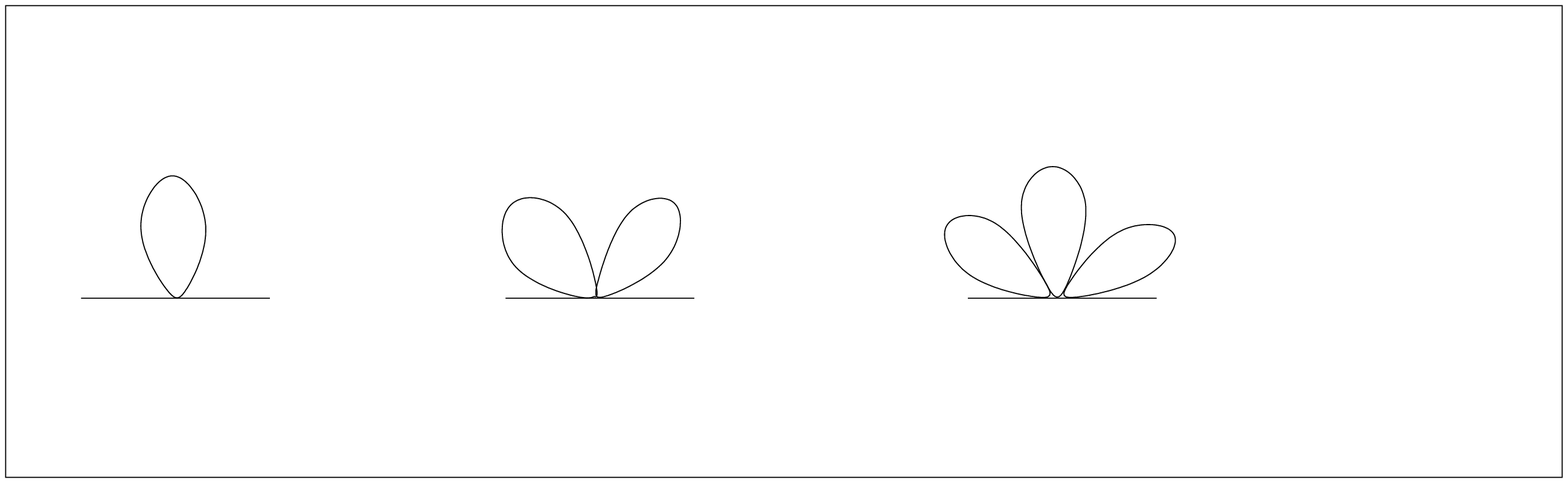,width=16cm, height=6cm}
\caption{The sum of all tadpole diagrams, which are finite at the external
  frequency $z\rightarrow 0$. They are not relevant for the glass transition
  problem.}
\end{figure}

\end{document}